\documentclass[trackchanges]{aastex701}
\usepackage{tabularx,rotating}
\usepackage{xcolor}
\usepackage{tablefootnote}
\usepackage{ulem} 
\usepackage{soul}
\usepackage{cancel}
\usepackage{wasysym} 
\usepackage{graphicx}
\usepackage{subcaption}
\usepackage{silence}
\newcommand{\sw}{{\it Swift}}
\newcommand{\cha}{{\it Chandra}}
\newcommand{\xmm}{XMM-{\it Newton}}

\newcommand{\Pa}{Paper~I} 
\newcommand{\Pb}{Paper~II}

\newcommand{\one}{MRC~B1754$-$597}
\newcommand{\two}{MRC~B1814$-$519}
\newcommand{\three}{MRC~B1817$-$391}
\newcommand{\four}{MRC~B1827$-$360}
\newcommand{\five}{MRC~B2032$-$350}
\shorttitle{A \sw~X-ray view of the SMS4 sample - III.}
\shortauthors{Maselli et al.}
\graphicspath{{./}{figures/}}
\begin{document}

\title{A \sw~X-ray view of the SMS4 sample - III: Deeper insight into previously undetected sources.}

\author[orcid=0000-0003-3760-1910, gname='Alessandro', sname='Maselli']{Alessandro Maselli} 
\affiliation{INAF-Osservatorio Astronomico di Roma, via Frascati 33, I-00078, Monte Porzio Catone (Roma), Italy}
\affiliation{ASI Space Science Data Center (SSDC), via del Politecnico snc, I-00133, Roma, Italy}
\email{alessandro.maselli@inaf.it}

\author[orcid=0000-0002-9478-1682]{William R. Forman}
\affiliation{Harvard-Smithsonian Center for Astrophysics, 60 Garden Street, Cambridge, MA-02138, USA}
\email{wforman@cfa.harvard.edu}

\author[orcid=0000-0003-2206-4243]{Christine Jones}
\affiliation{Harvard-Smithsonian Center for Astrophysics, 60 Garden Street, Cambridge, MA-02138, USA}
\email{cjones@cfa.harvard.edu}

\author[orcid=0000-0002-0765-0511]{Ralph P. Kraft}
\affiliation{Harvard-Smithsonian Center for Astrophysics, 60 Garden Street, Cambridge, MA-02138, USA}
\email{rkraft@cfa.harvard.edu}

\author[orcid=0000-0003-3613-4409]{Matteo Perri}
\affiliation{INAF-Osservatorio Astronomico di Roma, via Frascati 33, I-00078, Monte Porzio Catone (Roma), Italy}
\affiliation{ASI Space Science Data Center (SSDC), via del Politecnico snc, I-00133, Roma, Italy}
\email{matteo.perri@inaf.it}

\correspondingauthor{Alessandro Maselli}
\email{alessandro.maselli@inaf.it}

%\collaboration{all}{The Terra Mater collaboration}

%% Use the \collaboration command to identify collaborations. This command
%% takes an optional argument that is either a number or the word "all"
%% which tells the compiler how many of the authors above the command to
%% show. For example "\collaboration[all]{(DELVE Collaboration)}" wil include
%% all the authors above this command.
%%
%% Mark off the abstract in the ``abstract'' environment. 
\begin{abstract}

%MAX 250 WORDS
We update the X-ray information given in Maselli et~al.~(2024) for five bright radio sources in the SMS4 catalog, thanks to additional observations with the X-Ray Telescope (XRT) on board the Neil Gehrels \sw~Observatory (hereafter \sw) carried out through February~2026.
X-ray emission from \one, previously based only on data from the eROSITA-DE DR1 catalogs, is now detected also by \sw, which yields a more precise positional uncertainty. 
Thanks to $\sim$7~ks of additional exposure with \sw~we are now able to detect \three, a source for which no X-ray counterpart is found in eROSITA-DE DR1.  
Another source that is not detected in the eROSITA-DE DR1 catalogs, and that was not observed by \sw~earlier than 2025, is \four: with a $\sim$6~ks X-ray observation, the source is detected.
Finally, based on $\sim$8~ks of additional exposure, we detect X-ray emission for \five, a source that lies out of the DR1 footprint.
In contrast, additional exposure ($\sim$2~ks) just gives an upper limit on the X-ray emission of \two.
The analysis of the extent and the hardness ratio of the four detected X-ray sources suggests the presence of soft, diffuse X-ray emission as expected from galaxy group cores or hot galaxy coronae.
Using the positional uncertainty of the X-ray detections to constrain the search for counterparts at lower energies, we provide a new infrared/optical counterpart for \three~and confirm the counterparts previously reported in the literature for the three remaining sources. 

\end{abstract}

%% Keywords should appear after the \end{abstract} command. 
%% The AAS Journals now uses Unified Astronomy Thesaurus (UAT) concepts:
%% https://astrothesaurus.org
%% You will be asked to selected these concepts during the submission process
%% but this old "keyword" functionality is maintained in case authors want
%% to include these concepts in their preprints.
%%
%% You can use the \uat command to link your UAT concepts back its source.
%\keywords{\uat{Galaxies}{573} --- \uat{Cosmology}{343} --- \uat{High Energy astrophysics}{739} --- \uat{Interstellar medium}{847} --- \uat{Stellar astronomy}{1583} --- \uat{Solar physics}{1476}}

%% From the front matter, we move on to the body of the paper.
%% Sections are demarcated by \section and \subsection, respectively.
%% Observe the use of the LaTeX \label
%% command after the \subsection to give a symbolic KEY to the
%% subsection for cross-referencing in a \ref command.
%% You can use LaTeX's \ref and \label commands to keep track of
%% cross-references to sections, equations, tables, and figures.
%% That way, if you change the order of any elements, LaTeX will
%% automatically renumber them.

\section{Introduction} 
\label{sec:1}
To characterize bright radio-selected sources from SMS4 \citep{2006AJ....131..100B} in the X-rays, we initiated an X-ray survey \citep{2022ApJS..262...51M} using the Neil Gehrels Swift Observatory (hereafter \sw, \citealp{2004ApJ...611.1005G}) for all those objects that did not already have good quality X-ray images. 
In Maselli et al.~(2022; hereafter \Pa) we reported on \sw~observations with the X-Ray Telescope (XRT, \citealp{2005SSRv..120..165B}) of 31 radio sources selected from SMS4, while in Maselli et al.~(2024; hereafter \Pb), we described an additional sample of 17 targets, always observed by \sw.

In \Pb, six of these 17 sources remained undetected in the X-rays.
However, as of February~2026, several \sw~observations were collected for four of these six sources and also for \four, originally included in our proposal but not observed as of February 2025. 
Using all observations currently available for each source, we obtain four new X-ray detections and one remaining upper limit.

As in \Pa~and \Pb, we characterize the extent and the hardness ratio of the detected X-ray sources and use their positions to provide accurate locations for the central AGN engine of our sources, which were all classified as radio galaxies in \cite{2006AJ....131..114B}. 
We search the AllWISE \citep{2013wise.rept....1C}, CatWISE2020 \citep{2020ApJS..247...69E} and GSC~2.4.2 \citep{2008AJ....136..735L} catalogs for sources within the positional uncertainty of the four X-ray detected sources to identify their IR/optical counterparts. 
We require a detection in both the IR and the optical bands to establish a robust counterpart, and we find a candidate for all four X-ray detected sources.
We show their positions in the $W1$ AllWISE and $r$ DSS2 images and compare these with candidates previously given in the literature, mainly by \cite{2020PASA...37...18W,2020PASA...37...17W} (hereafter W20) in the infrared and by \cite{2006AJ....131..100B,2006AJ....131..114B} (hereafter BH06) in the optical.

Throughout this paper, we use CGS units, unless otherwise stated.
For consistency with \Pa~and \Pb, we assume a flat cosmology with $H_0 = 72$ km s$^{-1}$ Mpc$^{-1}$, $\Omega_M = 0.26$, and $\Omega_{\Lambda} = 0.74$ \citep{2009ApJS..180..306D}.

\begin{table*}
\footnotesize
\begin{center}
\caption{\small{Correspondences between the five SMS4 sources in our sample and G4Jy sources.}}
\label{tab:radio}
\begin{tabular}{cccc|ccccc}
\hline
Name   & S$_{178}$ & LAS      &    $z$                  & IAU Name            & G4Jy ID  &   l    &     b    & $\overline{S}_{181}$ \\ % & Morphology 
       &   (Jy)    & (arcsec) &                         &                     &          & (deg)  &   (deg)  &        (Jy)          \\  
 (1)   &   (2)     & (3)      &    (4)                  &   (5)               &   (6)    &  (7)   &    (8)   &         (9)          \\   
\hline
\one   &  20.0     &  21      &    (0.80)               &   J175906$-$594655  &  1453    & 333.77 & $-$17.02 & 25.16$\pm$0.04 \\ %s
\two   &  24.0     &  10      &    (0.48)               &   J181806$-$515801  &  1471    & 342.35 & $-$16.24 & 23.55$\pm$0.03 \\ %s
\three &  19.0     &  16      &    (0.91)               &   J182035$-$390925  &  1474    & 354.51 & $-$11.26 & 15.66$\pm$0.04 \\ %s
\four  &  33.0     &  10      &    (0.12)               &   J183059$-$360229  &  1487    & 358.29 & $-$11.78 & 30.15$\pm$0.04 \\ %s
\five  &  27.0     &  26      &    (0.56)               &   J203547$-$345403  &  1640    &   7.85 & $-$35.58 & 23.80$\pm$0.04 \\ %d
\hline                                                                                          
\end{tabular}
\end{center}
\tablecomments{The columns show (1) the name in SMS4, according to the MRC designation; (2) the extrapolated flux density $S_{178}$; (3) the largest angular size of the radio source at 843~MHz; (4) the redshift values taken from \cite{2006AJ....131..114B}, with photometric estimates in parentheses (for \four~a spectroscopic redshift $z$=0.078 has been recently obtained by \cite{2025PASA...42...85W}); (5) the International Astronomical Union (IAU) name in G4Jy, according to the GLEAM designation; (6) the G4Jy identifier; Galactic longitude (7) and latitude (8) of the G4Jy source; (9) the actual flux density $\overline{S}_{181}$ for G4Jy.\\
} 
\end{table*}

\section{Description of the newly observed SMS4 sources} 
\label{sec:2}
The five targets that complete our sample come from the list in \Pa~(see Table~8 therein) of 56 SMS4 bright radio sources with no \cha, \sw, or \xmm~observations as of September 2021.
Four of these sources were already investigated in the X-rays in \Pb, but remained undetected; an additional source, \four, had not been observed as of May 2024.

For each of these five SMS4 sources, a G4Jy source is found with an unambiguous match, with actual flux density values at 181~MHz in the range of 15-30~Jy, as derived from W20.
To compare the morphology of the radio source and the underlying X-ray emission, we use maps from the SUMSS survey to build radio flux density contours at 843~MHz and overlay these on the \sw-XRT X-ray maps generated as described in Section~\ref{sec:3}.
For each source, a single SUMSS component is found, as shown in Figure~\ref{fig:xrtmaps_soft_full}.
High-resolution maps at 5~GHz \citep{2006AJ....131..114B} from the Australian Telescope Compact Array (ATCA) reveal a more complex morphology, consistent with a FRII radio galaxy, for all but one source: the structure of \four~was considered not resolved enough for a classification. 
In Table~\ref{tab:radio}, we report for convenience all SMS4-G4Jy correspondences extracted from \Pb: quantities in columns 1--4 are taken from BH06, while those in columns 5--9 come from W20.

In BH06, for all five SMS4 sources, an optical counterpart was given using either the plates of the UK Schmidt Southern Sky Survey or R-band CCD images from the Anglo-Australian Telescope (AAT).
Among these, the optical counterpart of \four~has been investigated in greater detail by \cite{2025PASA...42...85W}, leading to a spectroscopic redshift $z$=0.078; for the other sources, only the photometric redshifts given by BH06 are currently available.
All redshift values are reported in Table~\ref{tab:radio}.

\begin{figure*}
\gridline{\fig{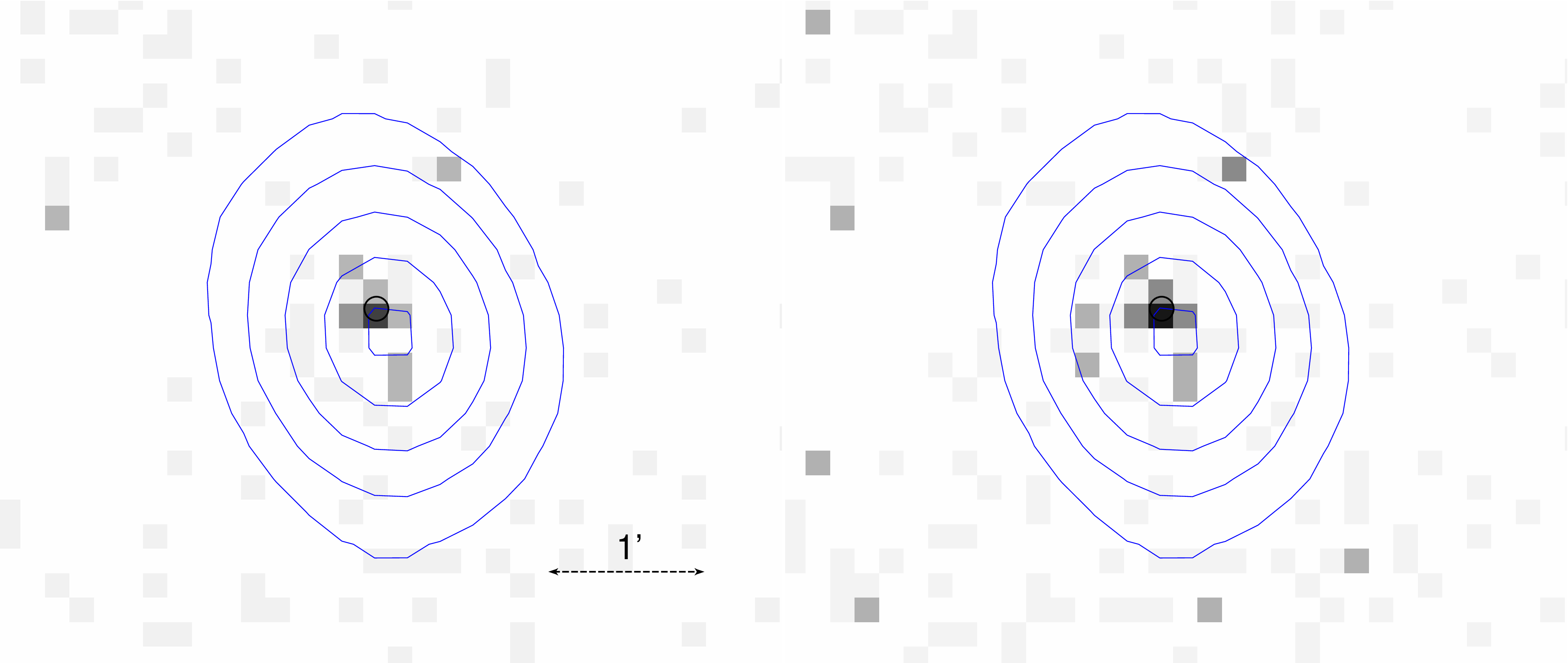}{0.70\textwidth}{\bf{ (a) \one~/ G4Jy 1453 }}}
\gridline{\fig{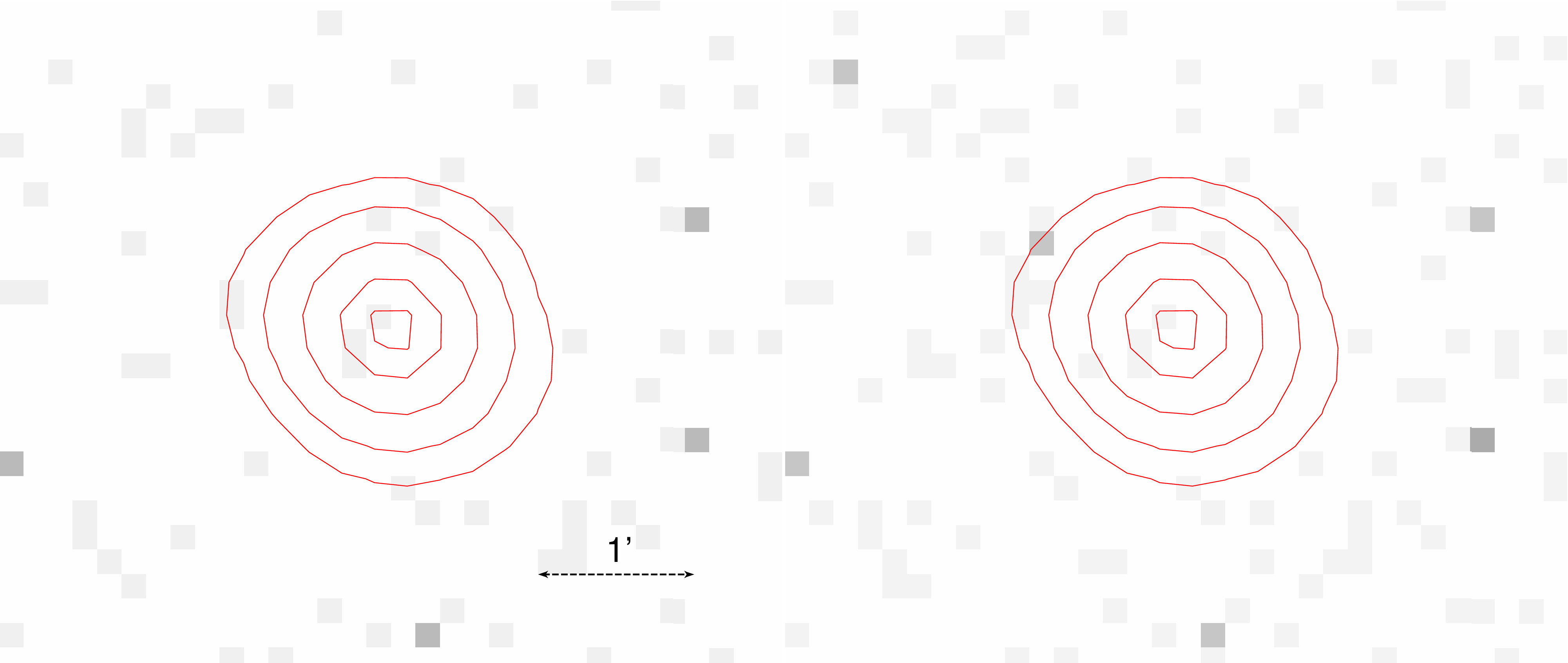}{0.70\textwidth}{\bf{ (b) \two~/ G4Jy 1471 }}}
\caption{\sw~X-ray maps (North is up, and East to the left) for \one~(a) and \two~(b). Two maps of the same field of view are given for each source: on the left panel, the map in the 0.5--4 keV band; on the right panel, in the 0.3--10 keV band. Black circles mark the positional uncertainty region for each X-ray detection, computed in the 0.3--10 keV band. All maps, centered at the positions of the SMS4 radio sources, are binned by 4x4 pixels. Radio flux density contours from the SUMSS survey, carried out at 843~MHz, overlay the X-ray maps and have been selected, for each source, to best display the shape of the radio emission. Following \Pa~and \Pb, blue contours distinguish sources detected in the X-rays by \sw~in our campaign from not detected sources, whose contours are marked in red.}
\label{fig:xrtmaps_soft_full}
\end{figure*}

\begin{figure*}
\ContinuedFloat
\gridline{\fig{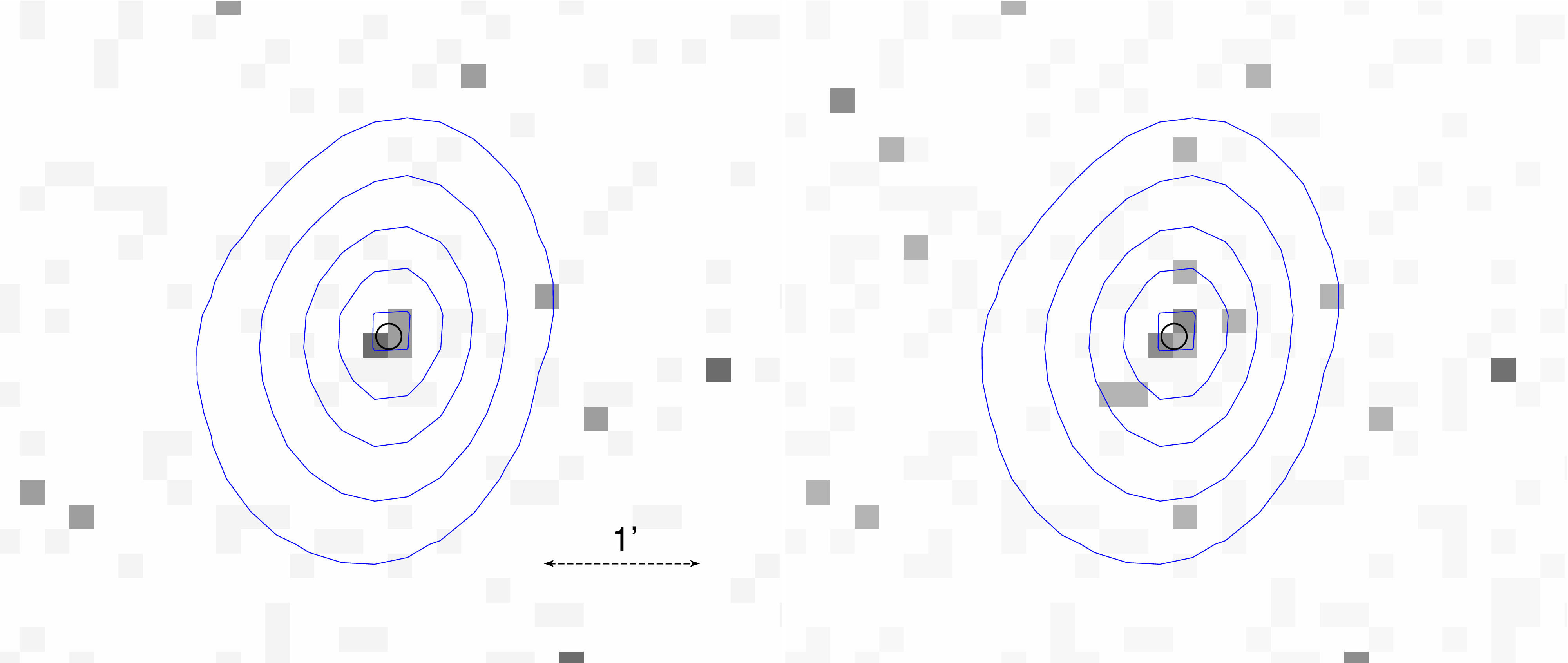}{0.70\textwidth}{\bf{(c) \three~/ G4Jy 1474}}}
\gridline{\fig{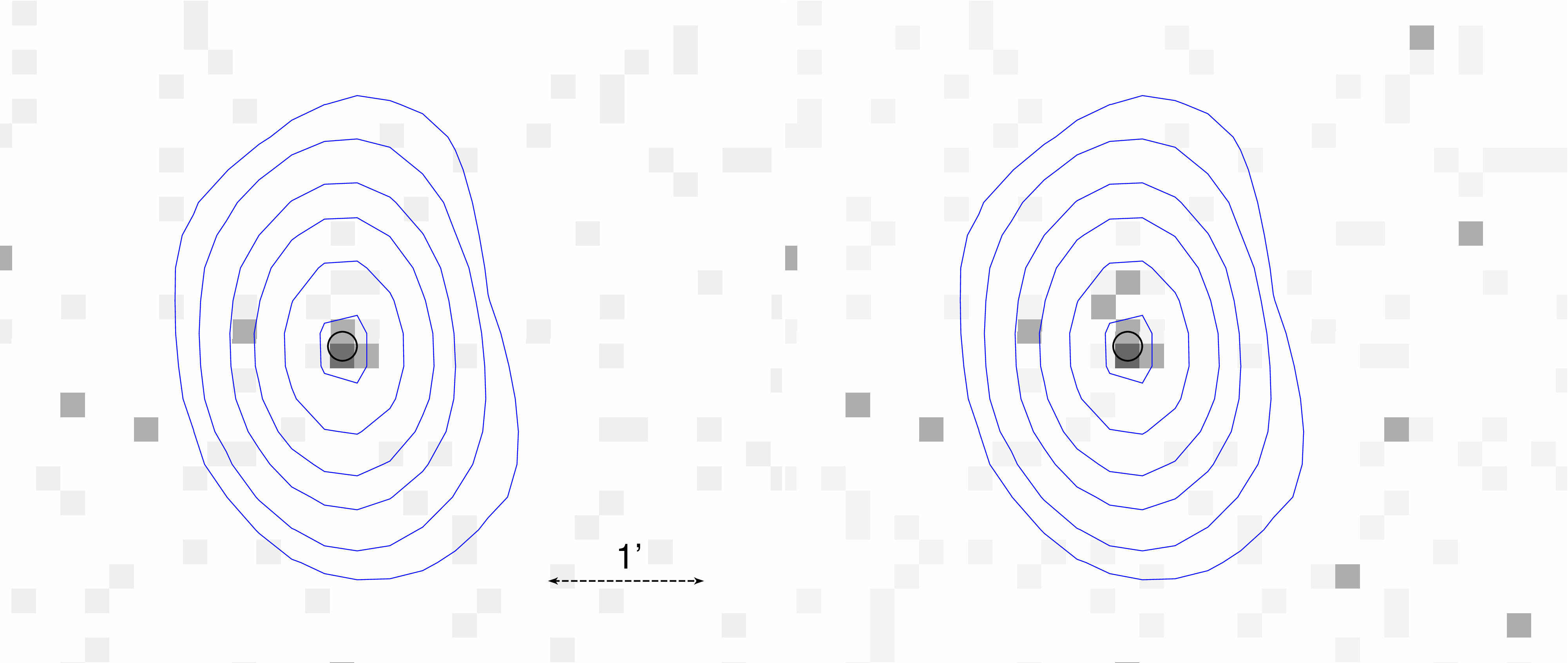}{0.70\textwidth}{\bf{(d) \four~/ G4Jy 1487}}}
\gridline{\fig{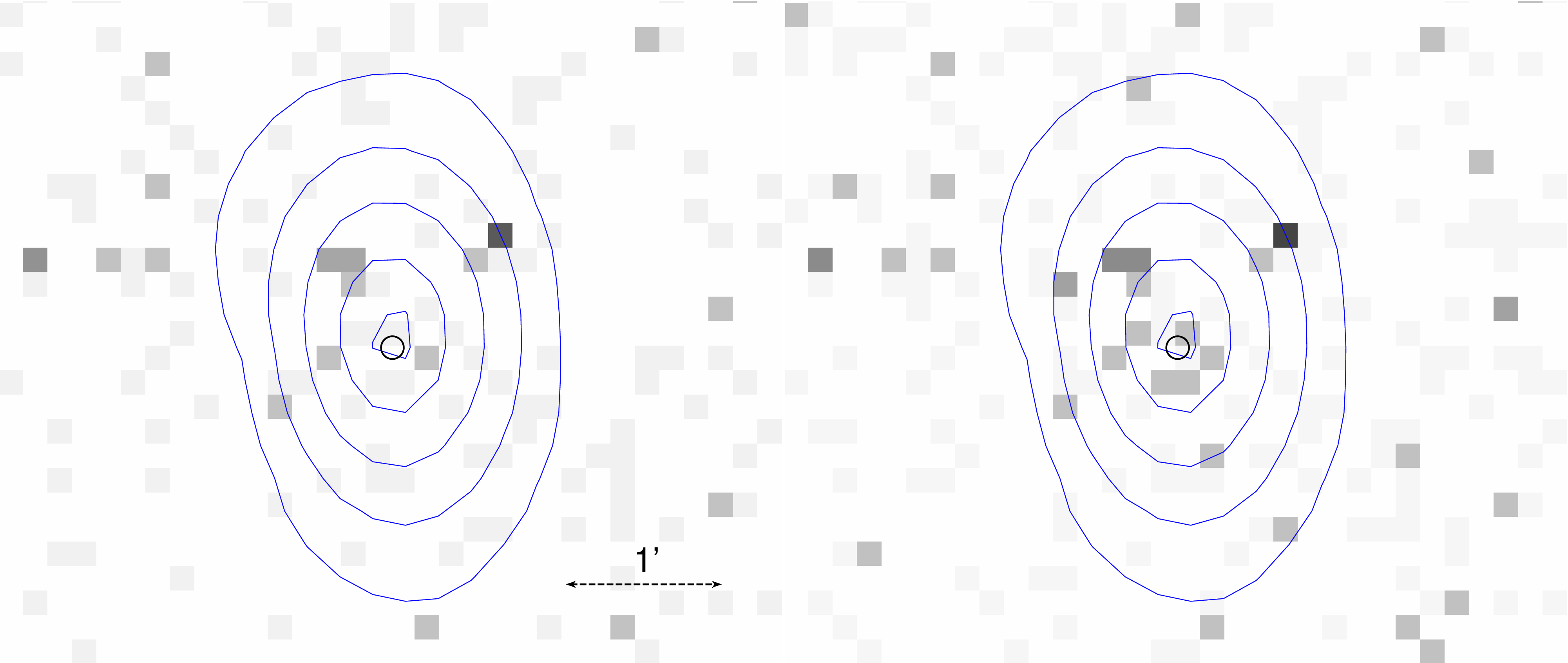}{0.70\textwidth}{\bf{(e) \five~/ G4Jy 1640}}}
\caption{(continued) \sw~X-ray maps (North is up, and East to the left) for \three~(c), \four~(d), and \five ~(e). Two maps of the same field of view are given for each source: on the left panel, the map in the 0.5--4 keV band; on the right panel, in the 0.3--10 keV band. Black circles mark the positional uncertainty region for each X-ray detection, computed in the 0.3--10 keV band. All maps, centered at the positions of the SMS4 radio sources, are binned by 4x4 pixels. Radio flux density contours from the SUMSS survey, carried out at 843~MHz, overlay the X-ray maps and have been selected, for each source, to best display the shape of the radio emission. Following \Pa~and \Pb, SUMMS contours are blue since all these three sources are detected in the X-ray band.}
\end{figure*}

\section{\sw-XRT data reduction and analysis} 
\label{sec:3}
In \Pb, the six sources that were reported as undetected in the X-rays were listed in Table~3.
However, several new observations have been carried out for four of these six undetected sources until February~2026; the two sources that were not observed again are MRC~B2041$-$604 and MRC~B2331$-$416.
These additional \sw~observations increased the exposure on the four observed sources: this increase is mild (2.2~ks) for \two, but is relevant for \one~(6.0~ks), for \three~(8.1~ks), and for \five~(8.0~ks).
Furthermore, new observations were performed for \four, which had not yet been observed, with a total exposure of 5.9~ks. 
As a result, we can update the X-ray information for five sources discussed in \Pb.

The procedures adopted to reduce X-ray data in the present analysis are analogous to those already described in both \Pa~and \Pb: below, we give essential details highlighting the updates and differences suitable for the present analysis.
X-ray data from the \sw-XRT, retrieved from the \sw~archive, are processed with the XRTDAS software package (v.3.7.1), developed at the Space Science Data Center (SSDC) of the Italian Space Agency (ASI) and distributed by the NASA High Energy Astrophysics Archive Research Center (HEASARC) within the HEASoft package (v.6.36).
Event files are calibrated and cleaned by applying standard filtering criteria with the {\sc xrtpipeline} task and using the latest calibration files available in the \sw~CALDB distributed by HEASARC. 
Events in the 0.3--10 keV energy range are used in our analysis, and exposure maps are created with {\sc xrtpipeline}.
Several observations are available for all sources: from each observation, event files and exposure maps are stacked with the {\sc xselect} and {\sc ximage} tools, respectively, to build a single event file and a single exposure map.
Fig.~\ref{fig:xrtmaps_soft_full} shows the X-ray maps in both the soft (0.5--4 keV, left side) and the full (0.3--10 keV, right side) bands for all five SMS4 sources included in our sample, built from the corresponding event files.

For each source, we simultaneously upload the event file and the exposure map within {\sc ximage} and compute the average background intensity over the whole detector with the {\sc background} command.
For all sources, the background value in the 0.3--10 keV band spans the range 5.9$\times$10$^{-4}$ ct~sqarcmin$^{-1}$~s$^{-1}$ (\five) to 9.5$\times$10$^{-4}$ ct~sqarcmin$^{-1}$~s$^{-1}$ (\one).
Taking into account the values of the Galactic latitudes reported in Table~\ref{tab:radio}, that are low ($\mid b \mid~< 20^\circ$) for most sources, the background values that we find are consistent with the distribution given in \cite{2011A&A...528A.122P}.

\begin{table*}
\tiny
\caption{Results from \sw-XRT observations.}
\label{tab:results}
\begin{center}
\begin{tabular}{ccccccccccc}
\hline
 SMS4 Name & R.A. (J2000) & Decl. (J2000)                &    $r_c$     &        Count Rate       & 3$\sigma$ Upper Limit &       $P$         & First Obs. & Latest Obs.& Obs. & Exposure \\
           & ($^{h~m~s}$) & ($^{\circ}$~\arcmin~\arcsec) &   (arcsec)   & (10$^{-3}$ ct s$^{-1}$) &   (ct s$^{-1}$)       &                   & (yy-mm-dd) & (yy-mm-dd) &      &   (s)    \\
 (1)       &  (2)         &             (3)              &    (4)       &         (5)             &          (6)          &       (7)         &    (8)     &    (9)     & (10) &   (11)   \\   
\hline                                                                                                                                 
\one       & 17 59 07.1   &        $-$59 46 50.8         &   4.6        &       5.3 $\pm$ 1.2     &        --             & $2.2\cdot10^{-16}$ & 22-05-22   & 25-02-04  &   13  &   6241  \\ % All J175907.08-594649.7  1.443
\three     & 18 20 35.4   &        $-$39 09 30.3         &   4.9        &       1.3 $\pm$ 0.5     &        --             & $1.3\cdot10^{-05}$ & 22-07-14   & 26-02-09  &   18  &  10806  \\ % Cat 
\four      & 18 30 58.8   &        $-$36 02 28.9         &   5.6        &       2.1 $\pm$ 0.8     &        --             & $6.7\cdot10^{-06}$ & 25-02-27   & 25-08-27  &    8  &   5942  \\ % All J183058.92-360230.7  0.502
\five      & 20 35 47.6   &        $-$34 54 08.9         &   4.4        &       0.8 $\pm$ 0.4     &        --             & $4.4\cdot10^{-04}$ & 22-09-02   & 24-03-13  &   10  &  13570  \\ % All J203547.67-345410.5  1.279	
\hline
\two       &   --         &             --               &    --        &            --           &  $2.4\cdot10^{-03}$   & $4.2\cdot10^{-01}$ & 23-02-12   & 25-11-09  &   11  &   5797  \\  % All J181806.87-515809.3	
\hline
\end{tabular}
\end{center}
\tablecomments{The columns show (1) the name of the corresponding SMS4 source; (2) the Right Ascension and (3) Declination of the X-ray detection; (4) the positional uncertainty $r_c$, at the 90\% confidence level; (5) the 0.3--10 keV count rate, with its uncertainty; (6) the 0.3--10 keV count rate 3$\sigma$ upper limit at the position of the radio coordinates; (7) the probability $P$ that the signal is a statistical fluctuation of the background; the dates of the first (8) and the latest (9) \sw~observation; (10) the number of stacked \sw~observations; (11) the total XRT exposure time at the coordinates of the SMS4 source. An horizontal line separates detected sources, with $P < 10^{-04}$, from the undetected source \two.}
\end{table*}

We perform a local source detection with the {\sc sosta} command within {\sc ximage}.
Based on our experience in the analysis of XRT images, we fix the side of the box to 14~pixels ($\sim33\arcsec$), a choice that maximizes the signal-to-noise ratio for faint sources.
For the background intensity, we fix it at the value previously obtained with the {\sc background} command.
For each X-ray image, the extraction box is centered exactly on the SMS4 coordinates.
For each source detection, {\sc sosta} provides the intensity of the source and its significance, in terms of the probability $P$ that the signal is a statistical fluctuation of the background.
We set $P^{\star} = 10^{-4}$ as a threshold to distinguish detections ($P < P^{\star}$) from non-detections, implying an order of magnitude tighter constraint with respect to the default in {\sc sosta}. 
As a result, considering the five SMS4 sources in our sample, our prescriptions lead to four new X-ray detections with \sw. 
To determine the position and its uncertainty (90\% confidence level) for all these detections, we use the {\sc xrtcentroid} task.
We give full observational details for our sample in Table~\ref{tab:results}.

We use WebPIMMS to convert the count rate of the XRT-detected sources into the X-ray unabsorbed flux in the 0.3--10 keV band.
We adopt two different spectral models, both absorbed by the Galactic hydrogen column density, to take into account two different emission mechanisms: a power law with photon index $\Gamma=2$ for AGN point-like sources and an APEC model (0.4 Solar abundance, $kT$=3 keV) for thermal emission from a diffuse source.

Then, we use the available redshift values to compute the X-ray luminosity in the 0.3--10 keV band.
For this purpose, for \four~we use the spectroscopic measurement recently obtained by \cite{2025PASA...42...85W}.
For \one~and \five, we use the photometric redshift of the optical counterparts reported in BH06, that we confirm (see Section~\ref{sec:4.3}).
Conversely, no luminosity was computed for \three~since the optical counterpart that we provide in Section~\ref{sec:4.3} is different from that proposed by BH06, and we have no available redshift. 
The results of our analysis are reported in Table~\ref{tab:pimms}.

\begin{table*} 
\begin{center}
\scriptsize
\caption{X-ray unabsorbed flux and luminosity for the \sw~detected SMS4 sources of our sample.}
\label{tab:pimms}
\begin{tabular}{ccccc|cc|cc}
                    \multicolumn{5}{c}{}                                   & \multicolumn{2}{c}{Power Law Model ($\Gamma=2$)} & \multicolumn{2}{c}{APEC Model (0.4 Solar, $kT$=3~keV)} \\  
 \hline
 Name  &  $z$   & $D_L$  &    ${n_{H,\,Gal}}$    &      XRT Count Rate     &      $S_{~0.3-10,unabs}$     &      $L_X$        &  $S_{~0.3-10,unabs}$     &        $L_X$      \\
       &        & (Mpc)  & (10$^{20}$ cm$^{-2}$) & (10$^{-3}$ ct s$^{-1}$) &   (erg cm$^{-2}$ s$^{-1}$)   &   (erg s$^{-1}$)  & (erg cm$^{-2}$ s$^{-1}$) &   (erg s$^{-1}$)  \\
  (1)  &   (2)  &  (3)   &         (4)           &             (5)         &              (6)             &        (7)        &       (8)                &        (9)        \\  
\hline
\one   & (0.80) & 4981.2 &        6.6            &            5.3          &      $2.2\cdot10^{-13}$      & $6.5\cdot10^{44}$ & $1.7\cdot10^{-13}$       & $5.2\cdot10^{44}$ \\
\three &  ...   &  ...   &        8.7            &            1.3          &      $5.7\cdot10^{-14}$      &      ...          & $4.5\cdot10^{-14}$       &       ...         \\ %NOT CONFIRMED BH06 CTP
\four  & 0.078  &  546.1 &        8.1            &            2.1          &      $9.0\cdot10^{-14}$      & $3.2\cdot10^{42}$ & $7.8\cdot10^{-14}$       & $2.8\cdot10^{42}$ \\
\five  & (0.56) & 3206.6 &        2.6            &            0.8          &      $3.0\cdot10^{-14}$      & $3.7\cdot10^{43}$ & $2.4\cdot10^{-14}$       & $3.0\cdot10^{43}$ \\
\hline
\end{tabular}
\end{center}
\tablecomments{The columns show (1) the name of the SMS4 source; (2) the redshift, with photometric values (in parentheses) from BH06 and a spectroscopic measurement from \cite{2025PASA...42...85W} (no redshift measurement is available for the optical counterpart that we provide in Section~\ref{sec:4.3} for \three, different from that given by BH06); (3) the corresponding luminosity distance $D_L$; (4) the Galactic hydrogen column density in the direction of the X-ray source; (5) the count rate detected by the \sw~XRT, as reported in Table~\ref{tab:results}; (6) the unabsorbed flux in the 0.3-10 keV band derived by WebPIMMS, assuming a power-law model with photon index $\Gamma=2$, with (7) the corresponding luminosity in the 0.3--10 keV band; (8) the same as column (6), but assuming an APEC model with a metal abundance corresponding to 0.4 Solar and a plasma temperature $kT=3$~keV, with (9) the corresponding luminosity in the 0.3--10 keV band.}
\end{table*}

\begin{figure*}[thbp]
\centerline{
\includegraphics[width=0.95\textwidth]{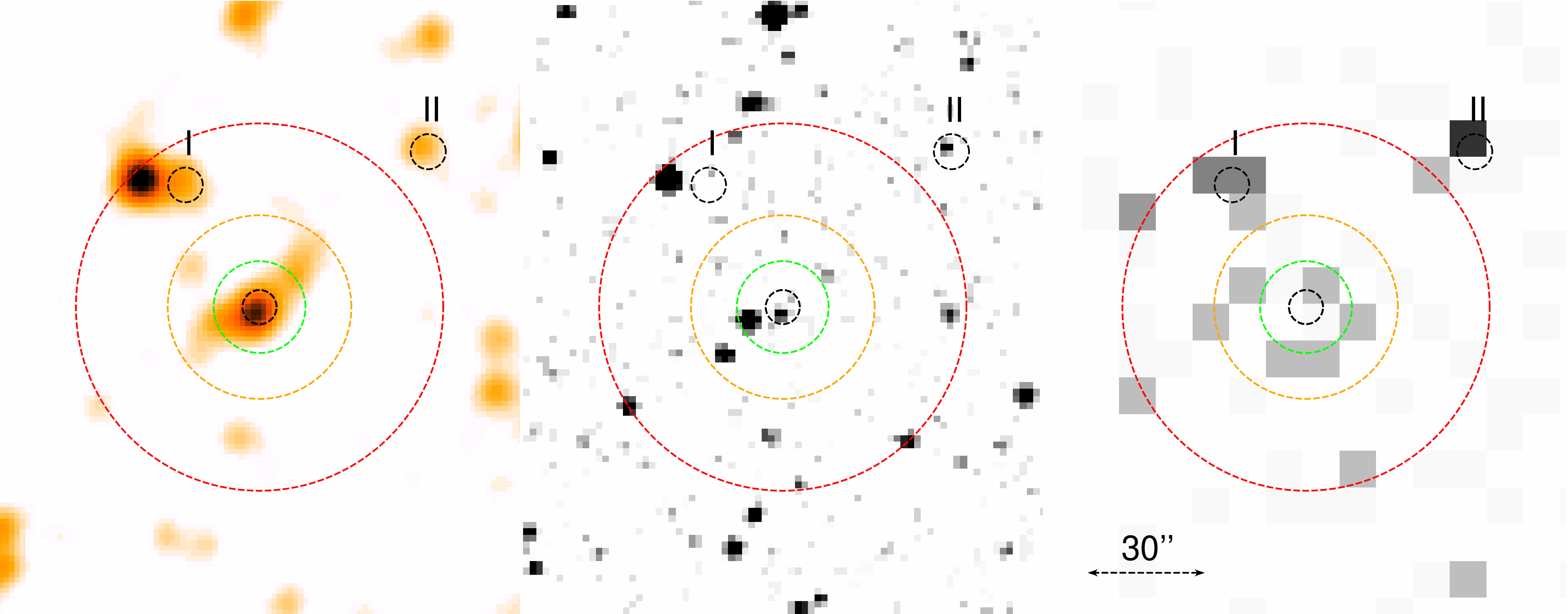}
}
\caption{Example of regions used to test sources for extent: Infrared, optical, and X-ray maps, matched in scale (N is up and E is to the left) and centered at the position of the X-ray detection for \five~(G4Jy 1640). The infrared map (left side) in the $W1$ filter (3.4 $\mu$m) is from AllWISE; the optical map (center) in the $r$ filter is from DSS2; the X-ray map in the 0.3--10 keV band is from \sw-XRT and is binned by 4x4 pixels. The black dotted circles mark the positional uncertainty of the examined X-ray detections, including the unrelated sources marked as $I$ and $II$ (see the text for further details). Three colored circles with radii of five (green), ten (orange), and twenty (red) pixels (1 pixel = 2\arcsec.36), respectively, overlay the maps to display the extraction regions (circle $C$ and annulus $A$) used to evaluate the extent of the X-ray emission, not only for \five~but for all the sources in Table~\ref{tab:er}.
}          
\label{fig:2032_neib}
\end{figure*}

\begin{table*} 
\begin{center}
\scriptsize
\caption{Extent of the X-ray emission.}
\label{tab:er}
\begin{tabular}{c|cccc|cccc}
                  &         \multicolumn{4}{c}{0.5--4 keV Band}         &     \multicolumn{4}{c}{0.3--10 keV Band}            \\  
\hline
Source Name       &     Background    &   $C$   & $A$  &      $ER$      &     Background    &   $C$   & $A$  &      $ER$      \\
                  & (10$^{-2}$ ct/px) &   (ct)  & (ct) &                & (10$^{-2}$ ct/px) &   (ct)  & (ct) &                \\
~~~~~~~~(1)       &        (2)        &   (3)   & (4)  &      (5)       &        (6)        &   (7)   & (8)  &      (9)       \\  
\hline                                                                                                                                 
\one              &       0.59        &    16   &  11  & $1.41\pm0.78$  &       0.99        &    19   &  14  & $1.37\pm0.68$  \\
\three            &       0.80        &     6   &  12  & $0.53\pm0.36$  &       1.27        &     7   &  18  & $0.39\pm0.24$  \\      
\four             &       0.53        &     9   &   8  & $1.07\pm0.75$  &       0.81        &     8   &  10  & $0.84\pm0.55$  \\      
\five             &       0.78        &     2   &  11  & $0.22\pm0.21$  &       1.36        &     6   &  16  & $0.37\pm0.25 $ \\ % excluding I
\hline                                              
\end{tabular}
\end{center}
\tablecomments{The columns show (1) the name of the SMS4 source; (2) and (6) the measured number of background counts, per pixel (1~pixel = 2\arcsec.36~on a side); (3) and (7) the background-corrected number of counts $C$ within a circle with a radius of 5~pixels; (4) and (8) the background-corrected number of counts $A$ within an annulus with inner and outer radii of 10 and 20 pixels, respectively; (5) and (9) the extent ratio $ER$, with its 1$\sigma$ uncertainty. The significance of source extent can be evaluated by comparing the measured $ER$ value (columns 5 and 9) to that for the model PSF that, for an unresolved source, yields $ER=5.73$ in both energy bands. For \five, we excluded the events corresponding to the unrelated source, labeled \textit{I} within the annulus, in all our analyses (see Figure~\ref{fig:2032_neib}, right panel).}
\end{table*}

\subsection{Extent of the X-ray Emission}
\label{sec:3.1}
For each X-ray detected source in Table~\ref{tab:results}, we compare the radial distribution of the  events with that expected from a point-like source to evaluate the extent of X-ray emission.
Such analysis is based on the analytic PSF model for the \sw-XRT telescope, given by the sum of a Gaussian and a King law as derived by the XRT Calibration Team (\url{https://swift.gsfc.nasa.gov/caldb/docs/xrt/SWIFT-XRT-CALDB-10_v01.pdf}).
We perform a ratio test based on a comparison of the number of events in a circle $C$ and a surrounding annulus $A$, both centered on the coordinates of the X-ray centroid given in Table~\ref{tab:results}.
The effectiveness of this method was confirmed in \Pb~using the point-like source MRK~876, a Seyfert~I galaxy.

Following \Pa~and \Pb, we use as extraction regions for our analysis a circle $C$ with a radius of 5~pixels ($\sim12\arcsec$) for the source and an annulus $A$ with inner and outer radii of 10 and 20 pixels, respectively, for the background; both regions are shown for example in Figure~\ref{fig:2032_neib} for the source \five.
After extracting the counts from the circles $C$ and the annuli $A$, we correct them for the background contribution, computed as reported in Section~\ref{sec:3}.  
Our analysis is carried out not only in the 0.3--10.0 keV band but also in the 0.5--4 keV band, to focus more on thermal emission that is appropriate to groups and clusters.  
The fraction $ER=C/A$ expected from a point-like source, computed from the analytic PSF model, is equal to 5.73 in both energy bands.
Thus, a deviation from this value reveals that the distribution of events differs from that expected from a point-like source, making $ER$ a marker of extended X-ray emission.

While verifying the presence of unrelated sources within the extraction regions, for \five~we find evidence of an excess, marked as $I$ in Figure~\ref{fig:2032_neib}, definitely within the annulus; another excess, external to the annulus but in its proximity, is marked as $II$.
As shown by comparing Figure~\ref{fig:xrtmaps_soft_full} (e panels) and Figure~\ref{fig:2032_neib}, both $I$ and $II$ are covered by radio flux density contours of \five. 
As performed for any other X-ray detection (see Section~\ref{sec:3}) we establish for $I$ the centroid (R.A.~(J2000)~=~20$^h$~35$^m$~49$^s$.1, Dec~(J2000)~=~$-$34$^{\circ}$~53\arcmin~37\arcsec.5; $r_I = 4\arcsec.5$), resulting in an angular separation of 36\arcsec.7 from the X-ray detection for \five.
Considering for $I$ a positional uncertainty $r_I = 4\arcsec.5$, we find that it matches the catalogued infrared source CatWISE J203549.17$-$345335.1, that is well distinguished in the AllWISE W1 map shown in Figure~\ref{fig:2032_neib}.
For this reason, we consider that the $I$ excess is due to an unrelated source.
The same analysis, carried out also for $II$ (R.A.~(J2000)~=~20$^h$~35$^m$~44$^s$.0, Dec~(J2000)~=~$-$34$^{\circ}$~53\arcmin~29\arcsec.0; $r_{II} = 4\arcsec.5$) leads to identify another distinct source emitting in the X-rays, at 58\arcsec.9 from \five. 
For $II$, an infrared (CatWISE J203544.19$-$345327.6.) and an optical (GSC\,2.4.2 SCMJ107625) counterpart are found, and both are well distinguished in Figure~\ref{fig:2032_neib}.

The $ER$ values of the X-ray detected sources, for the 0.5--4~keV and 0.3--10~keV bands, are reported in Table~\ref{tab:er}.
In particular, for \five, we exclude the contribution from the $I$ source, replacing the number of events detected in a 8x8 pixel box including the $I$ coordinates with that expected from the background.
As a result, we replace eight and ten events accumulated in the 0.5--4~keV and in the 0.3--10~keV bands, respectively, with one background event expected in both bands.
According to the results shown in Table~\ref{tab:er}, none of the sources is consistent with being point-like.

\begin{table*} 
\begin{center}
\scriptsize
\caption{Hardness ratio of the X-ray detections.}
\label{tab:hr}
\begin{tabular}{c|ccc|ccc}
\hline
Name   & $S$ (ct) &  $H$ (ct) &       $HR$      & $S'$ (ct) & $H'$ (ct) &      $HR'$     \\
  (1)  &  (2)     &    (3)    &       (4)       &    (5)    &    (6)    &       (7)      \\  
\hline                                                                                           
\one   &    23    &     6     &  $-0.59\pm0.14$ &    16     &      13   & $-0.10\pm0.24$ \\
\three &    15    &     1     &  $-0.88\pm0.09$ &     8     &       4   & $-0.33\pm0.32$ \\
\four  &    13    &     1     &  $-0.86\pm0.10$ &    11     &       1   & $-0.83\pm0.14$ \\
\five  &    10    &     7     &  $-0.18\pm0.30$ &    10     &       5   & $-0.33\pm0.27$ \\
\hline
\end{tabular}
\end{center}
\tablecomments{The columns show (1) the name of the SMS4 source; (2) the measured number of counts $S$ in the soft (0.3--3 keV) band within a circle of 10~pixels (23\arcsec.6) centered at the coordinates of the X-ray centroid, reported in Table~\ref{tab:results}; (3) the measured number of counts $H$ within the same circle, but in the hard (3--10 keV) band; (4) the corresponding hardness ratio $HR$, with its 1$\sigma$ uncertainty; (5) the same as column (2), but in the 0.5--2 keV band; (6) the same as column (3), but in the 2--7 keV band; (7) the corresponding hardness ratio $HR'$, with its 1$\sigma$ uncertainty.}
\end{table*}

\subsection{Hardness Ratio} 
\label{sec:3.2}
We use the {\sc counts} command within {\sc ximage} to extract the counts within a circle with a radius of 10~pixels (23\arcsec.6) centered at the position of our X-ray detections.
Following \Pa~and \Pb, we distinguish the number of counts in the soft ($S$ in the 0.3--3 keV) and hard ($H$ in the 3--10 keV) X-ray bands to compute the hardness ratio as $HR = (H - S) / (H + S)$.
As shown in Table~\ref{tab:hr}, despite some differences in the $HR$ values, no source shows evidence of hard emission.
Only \five~shows a relatively higher number of counts $H$ in the hard band, but in any case lower than $S$. 
For this source, we emphasize that the source $I$, being external to the circle with a radius of 10 pixels, does not affect the hardness ratio analysis.

\cite{2012A&A...547A..57T} used \sw~observations to distinguish AGN from galaxy group or galaxy cluster emission, based on the hardness ratio of the detected X-ray emission.
In their Figure~6, these authors show that, in a redshift {\it vs} HR plot, the region $HR>-0.2$ is mainly populated by AGNs, while the region $HR<-0.4$ characterizes thermal emission from groups and clusters.  
To compare our sources with that study, we repeat our analysis in the same energy bands adopted therein, that is 0.5--2 keV and 2--7 keV for the soft and the hard bands, respectively.
As shown in Table~\ref{tab:hr}, \four~is the only source which provides robust evidence of emission consistent with a group or a cluster.
For the remaining detected sources, taking into account 1$\sigma$ uncertainties, despite negative HR values also in the modified energy bands, conclusions are less stringent, with HR values consistent with the intersection region $-0.4 < HR < -0.2$.

\subsection{Comparison with other X-ray Catalogs} 
\label{sec:3.3}
The Living \sw-XRT Point Source catalog (LSXPS, \citealp{2023MNRAS.518..174E}) is a dynamic catalog that is updated almost in real time, as soon as new \sw~observations are archived; in addition to the whole catalog, two subsets ({\it clean} and {\it ultra-clean}) are found in the LSXPS. 
In January 2024, data from the first eROSITA-DE All-Sky Survey (eRASS1) were released \citep{2024arXiv240117274M}: three catalogs ({\it Main, Hard, and Supplementary}) are found in the First Data Release (DR1).
Here we compare the results of our \sw~campaign with the LSXPS and the eROSITA-DE DR1 catalogs.

\begin{table*} 
\begin{center}
\scriptsize
\caption{Positional match of new X-ray detections with other X-ray catalogs.}
\label{tab:xcat}
\begin{tabular}{c|ccc|cc|cc}
       \multicolumn{4}{c}{}                  &  \multicolumn{2}{c}{LSXPS}  &  \multicolumn{2}{c}{DR1} \\   
\hline
SMS4 Source & $r_{c}$  & $r_{l}$  & $r_{e}$  & $d_{cl}$ & $r_{cl}$ &       $d_{ce}$       & $r_{ce}$ \\
            & (arcsec) & (arcsec) & (arcsec) & (arcsec) & (arcsec) &       (arcsec)       & (arcsec) \\
   (1)      &   (2)    &   (3)    &   (4)    &   (5)    &   (6)    &         (7)          &   (8)    \\  
\hline                                                                                                                                 
\one        &   4.6    &   7.9    &   8.9    &    1.9   &   9.1    &         8.2          &  10.0    \\
\three      &   4.9    &   ...    &   ...    &    ...   &   ...    &         ...          &   ...    \\
\four       &   5.6    &   5.7    &   ...    &    2.7   &   8.0    &         ...          &   ...    \\
\five       &   4.4    &   ...    & No Data  &    ...   &   ...    &       No Data        & No Data  \\
\hline
\end{tabular}
\end{center}
\tablecomments{The columns show (1) the name of the SMS4 source; (2) the positional uncertainty (90\% c.l.) of the \sw~detection from our campaign $r_{c}$; (3) the positional uncertainty (90\% c.l.) of the LSXPS source $r_{l}$; (4) the positional uncertainty (90\% c.l.) of the eROSITA-DE DR1 source $r_{e}$; (5) the angular separation $d_{cl}$ between our detection and the LSXPS source; (6) the matching radius $r_{cl}$ between our detection and the LSXPS source; (7) the angular separation $d_{ce}$ between our detection and the eROSITA-DE DR1 source; (8) the matching radius $r_{ce}$ between our detection and the eROSITA-DE DR1 source. Three dots indicate sources that are undetected in a catalog; \five~is out of the eROSITA-DE footprint.}
\end{table*}

Searching for X-ray sources in the LSXPS (\url{https://www.swift.ac.uk/LSXPS/}) within 3\arcmin~from the coordinates of our X-ray detections, we find one LSXPS source only for \one~and \four; when found, the LSXPS source belongs to the {\it ultra-clean} sample, with no other sources in the whole catalog.
In both cases we compute the angular separations $d_{cl}$ between our detections and the LSXPS sources, as well as the matching radius $r_{cl}=(r_{c}^2+r_{l}^2)^{1/2}$, where $r_{c}$ and $r_{l}$ are the positional uncertainties of corresponding sources, respectively.
As shown in Table~\ref{tab:xcat}, we find that $d_{cl}<r_{cl}$ for both sources.
No values are given in Table~\ref{tab:xcat} for \three~and \five, that are undetected in LSXPS.

In the eROSITA data archive (eRODat, \url{https://erosita.mpe.mpg.de/dr1/erodat/}), a 1eRASS counterpart in the {\it Main} (0.2--2.3 keV) catalog, is found only for \one.
This 1eRASS source was already given as X-ray counterpart in \Pb, but now we can compute the angular separation $d_{ce}$ from the \sw-XRT detection, and find that it is lower than the matching radius $r_{ce}=(r_{c}^2+r_{e}^2)^{1/2}$, as shown in Table~\ref{tab:xcat}. 
No values are given in Table~\ref{tab:xcat} for \three~and \four, both undetected in eROSITA-DE DR1, and also for \five, that is out of the eROSITA-DE footprint.
For all the \sw-XRT detections, nothing relevant is found neither in the {\it Hard} (2.3--5.0 keV) nor in the {\it Supplementary} catalog, that includes lower confidence detections.

Finally, as expected, we find neither an LSXPS nor an eROSITA source for \two, the SMS4 source undetected by \sw~(Table~\ref{tab:results}).

\begin{figure*}
\gridline{\fig{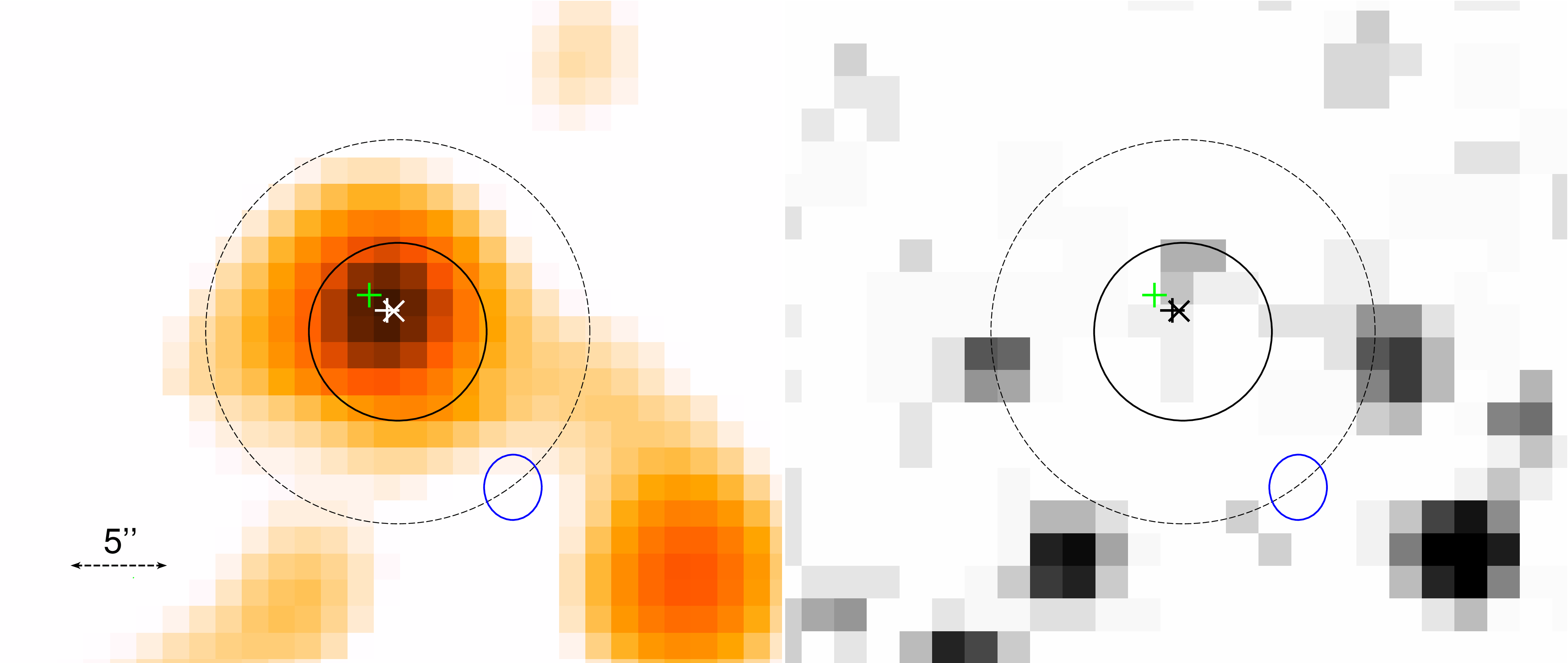}{0.95\textwidth}{\bf{( \one~/ G4Jy 1453 )}}}
\gridline{\fig{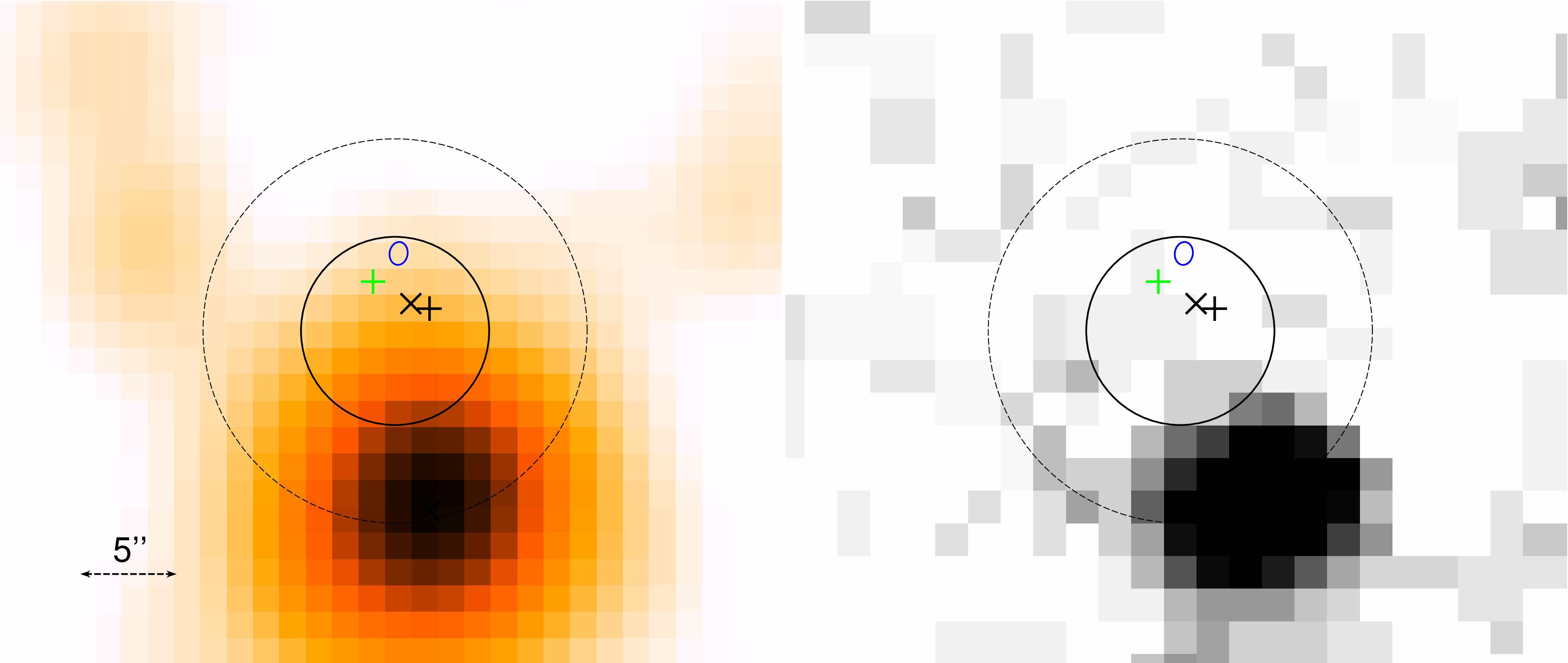}{0.95\textwidth}{\bf{( \three~/ G4Jy 1474 )}}}
\caption{Infrared and optical maps, (matched in scale for each source; N is up and E is to the left), of the four SMS4 sources detected by \sw-XRT in our sample, centered at the coordinates of the X-ray detections. Infrared maps (left side of each panel) in the $W1$ filter (3.4 $\mu$m) are from AllWISE, while optical maps (right side) in the $r$ filter are from DSS2. Blue ellipses mark the G4Jy positional uncertainty regions. Black solid circles mark the positional uncertainty of the X-ray sources, given at 90\% confidence level, while dashed circles mark the radio/X-ray positional match region at 95\% confidence level. Crosses (x) and plus signs (+) mark infrared sources from AllWISE and optical sources from GSC~2.4.2. Black or white are equally used for AllWISE and GSC~2.4.2 sources to improve the visibility with respect to the map in the background. Green is used for AllWISE counterparts also associated by W20, and for the optical counterparts suggested by BH06.}
\label{fig:iroptcand}
\end{figure*}

\setcounter{figure}{2}
\begin{figure*}
\gridline{\fig{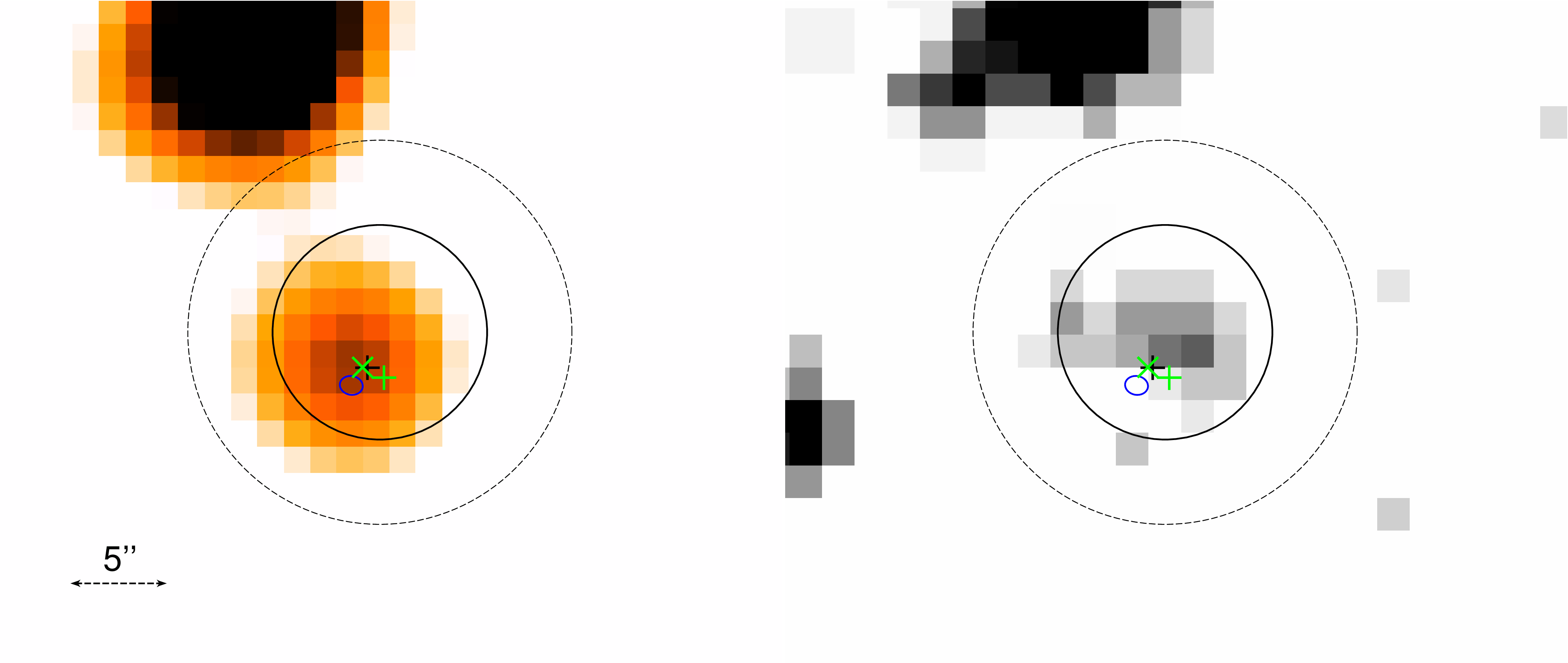}{0.95\textwidth}{\bf{( \four~/ G4Jy 1487 )}}}
\gridline{\fig{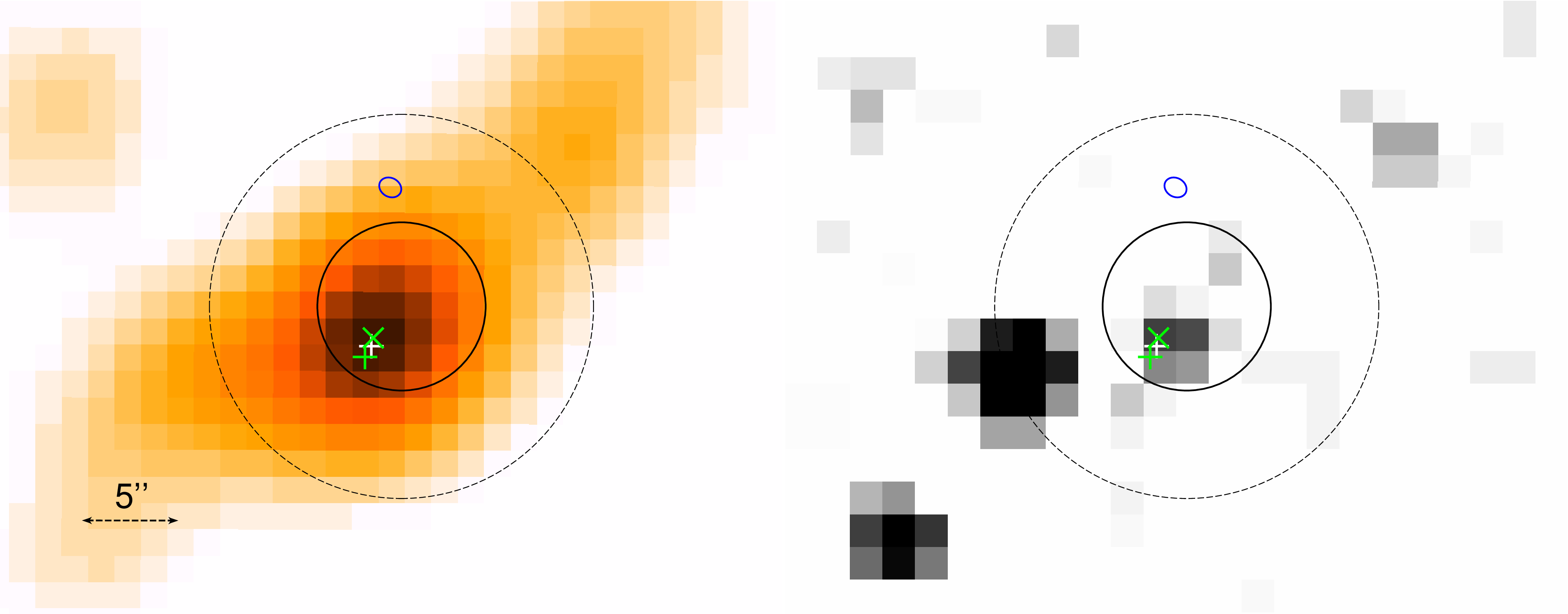}{0.95\textwidth}{\bf{( \five~/ G4Jy 1640 )}}}
\caption{Continued.}
\end{figure*}

\section{Multi-frequency analysis} 
\label{sec:4}
In this Section, we first discuss the matching of the SMS4 radio sources with the detected X-ray sources.
Using comparable values for the positional uncertainties of both radio and X-ray sources, we showed in \Pa~that the probability of chance coincidence between a random X-ray detection and an SMS4 source is negligible. 
We then discuss the WISE and optical identifications, mainly utilizing the X-ray position: the association with infrared or optical sources is allowed only if their coordinates lie within the corresponding X-ray positional uncertainty.

As reported in \Pa~(Table~8), BH06 did not give any morphological classification for \four, while the three remaining SMS4 sources detected in X-rays, by \sw, were classified as radio galaxies with an FRII morphology.
In the optical band, emission from the AGN itself, but likely also from the host galaxy, can be detected.
In type~II AGNs, the nuclear optical emission might be absorbed and obscured, but the infrared is more likely to reach the observer.
In our efforts to localize the core of the AGN, following the same criterion already used in \Pa~and \Pb, with our multifrequency analysis, we require a detection in both the infrared and optical bands.

We use TOPCAT to cross-match our list of X-ray detections with sources from selected radio, infrared, and optical catalogs.
To display the match of our X-ray detections with infrared and optical counterparts, we retrieve infrared maps in the $W1$~filter (3.4~$\mu$m) from the AllWISE Data Release \citep{2014yCat.2328....0C} Images Atlas, and optical maps in the $r$~filter (0.62~$\mu$m) from the Space Telescope Science Institute (STScI) $2^{nd}$ Digitized Sky Survey (DSS2) using the SkyView Virtual Observatory.
These maps are shown in Fig.~\ref{fig:iroptcand}.

\subsection{Associating X-ray Sources with Radio Sources}
\label{sec:4.1}
To reliably associate the four X-ray detected sources listed in Table~\ref{tab:results} with corresponding radio sources, we match the source positions at these two energy ranges using G4Jy as the main matching radio catalog for the sources.
Following W20, the typical rms positional uncertainties for G4Jy sources are $\sigma_{\alpha,S}$ $\approx$ 1\arcsec.5, $\sigma_{\delta,S}$ $\approx$ 1\arcsec.7, when their brightness-weighted centroids were computed after a cross-correlation with SUMSS data, while they are $\sigma_{\alpha,N}$ $\approx$ 0\arcsec.5, $\sigma_{\delta,N}$ $\approx$ 0\arcsec.6, when the same operation was based on NVSS data.

To match the radio and X-ray positions, we conservatively use a circular confidence region sufficiently large to ensure, with high probability, that the true radio source location lies within the X-ray positional uncertainty region.
For this circular region, we use a radius $r_X$ corresponding to the error radius, at 95\% confidence level, for sources at the X-ray limit of sensitivity, to which we add the largest uncertainty $r_r$ for the radio position, at the same confidence level as for the X-rays.
For the X-ray band, we establish $r_X=6\arcsec.7$, that is the positional uncertainty of \four~(see Table~\ref{tab:results}) multiplied by a factor 1.19 derived from the Normal distribution.
For the radio band, we take the larger of $\sigma_{\alpha,S}$ and $\sigma_{\delta,S}$, and multiply it by a factor of 1.96, derived from the Normal distribution, obtaining $r_r=3\arcsec.3$.

Using a circle with $R_{95}=r_r+r_X=$10$\arcsec.0$, we verify that the combined 95\% confidence positional uncertainty (the dashed black circles in Fig.~\ref{fig:iroptcand}) overlaps the corresponding G4Jy positional uncertainty for \one~and totally includes it for all the remaining \sw-XRT detections.
In particular we note that, for \three~and \four, the G4Jy ellipse is totally included right within the X-ray error region (the solid black circles in Fig.~\ref{fig:iroptcand}).

\begin{table*} 
\begin{center}
\scriptsize
\caption{Candidate counterparts for X-ray detected SMS4 sources.}
\label{tab:ir_opt}
\begin{tabular}{c|ccccc}
\hline
Name   &        Infrared Source         & $W1$ (mag) & $W1-W2$ (mag) & Optical Source & $\Delta$ (arcsec) \\
%      &                                &   (mag)    &    (mag)      &                &    (arcsec)       \\
 (1)   &              (2)               &    (3)     &     (4)       &       (5)      &       (6)         \\  
\hline                                                                                                                                            
\one   & AllWISE J175907.08$-$594649.7  &   14.49    &      1.44     &   S7FD049277   &       0.35        \\                   
\three & CatWISE J182035.30$-$390928.7  &   15.40    &      0.99     &   S9Q4111853   &       0.98        \\
\four  & AllWISE J183058.92$-$360230.7* &   12.74    &      0.50     &   S9P3007448   &       0.25        \\ 
\five  & AllWISE J203547.67$-$345410.5* &   14.46    &      1.28     &   SCMJ107169   &       0.44        \\ 
\hline                                                                                       
\end{tabular}
\end{center}
\tablecomments{\scriptsize{The columns show (1) the name of the SMS4 source; (2) the AllWISE \citep{2014yCat.2328....0C} or CatWISE \citep{2021ApJS..253....8M} infrared source identification; (3) the WISE magnitude in the $W1$ filter; (4) the WISE $W1-W2$ infrared color; (5) the GSC\,2.4.2 \citep{2021yCat.1353....0L} optical source identification; (6) the angular separation $\Delta$ between the given infrared and optical sources.\\
$*$: this infrared source was previously associated by W20. \\
}}
\end{table*}

\subsection{Cross Matches with Infrared and Optical Catalogs}
\label{sec:4.2}
In the infrared band, in our cross match with the X-ray detected sources we adopt AllWISE, and the typical positional uncertainties for the AllWISE sources are $\simeq$0\arcsec.06. 
For \three, since no counterpart is found in AllWISE, we search and find it in CatWISE2020, with comparable positional uncertainty.
As in \Pa, we take into account the infrared color of our WISE candidates, comparing $W1-W2$ with the threshold ($W1-W2\geq0.8$ mag) established by \cite{2012ApJ...753...30S} in their simple criterion for selecting AGNs.
In the optical band, the catalog that we use for source identification is the 2$^{nd}$ Generation Guide Star Catalog (GSC~2.4.2) by \cite{2021yCat.1353....0L}.
The astrometric information for this version\footnote{https://cdsarc.cds.unistra.fr/ftp/I/353/ReadMe} of the GSC catalog is updated to Gaia DR2, leading to very low positional uncertainties, of the order of a fraction of milliarcseconds, as in the cases of \three~(0\arcsec.0002) and \four~(0\arcsec.0007); higher values are found for \one~(0\arcsec.4) and \five~(0\arcsec.03).

For all the four SMS4 sources detected by \sw, we find one candidate counterpart, detected in both the infrared and the optical bands.
These candidates are unique, and lie within the boundaries of the X-ray positional uncertainty.
This occurrence qualifies all the SMS4 sources as class~A sources, as established in \Pa~and \Pb.
However, while we are confident in the identification of the counterparts, \four~does not have IR AGN colors. 
Thus, following \Pb, we define \four~as class~A2, with the remaining sources as class~A1. 
In particular, the \sw~X-ray detection allows an upgrade from B to A1 in the classification for \one, that was previously based on the larger positional uncertainty of the eROSITA source in DR1, still confirming the candidate previously given in \Pb.

All our candidate counterparts are listed in Table~\ref{tab:ir_opt}:
as shown in column (6), the angular distance between the infrared and optical sources is always lower than 1\arcsec, attesting to the fact that they can be both attributed to the same astrophysical source.

\subsection{Comparison with Earlier Studies}
\label{sec:4.3}
W20 associated an infrared counterpart from AllWISE to two (\four~and \five) of the four SMS4 sources for which X-ray emission was detected: with our approach, we confirm these two infrared counterparts.
Moreover, with a tighter constraint on the X-ray positional uncertainty of the X-ray source given by the \sw-XRT detection with respect to eROSITA-DE DR1, we are now able to remove the uncertainty about the counterpart to \one, previously raised in \Pb, and confirm the candidate that was already addressed as the most probable. 
Finally, we provide, for the first time, an infrared counterpart for \three, which is reported only in CatWISE2020.
As shown in Figure~\ref{fig:iroptcand}, this source is found at only 10\arcsec.5 from the much brighter source AllWISE~J182035.22$-$390939.6.

Optical counterparts were associated by BH06 to each of the four X-ray detected sources, using either the plates from the UK Schmidt Southern Sky Survey or R-band CCD images from a dedicated campaign at the AAT.
As shown in Table~\ref{tab:opt}, we find agreement between the coordinates reported in the GSC~2.4.2 catalog and those reported by BH06, with angular distances in most cases less than 1\arcsec.2, leading us to conclude that both optical sources refer to the same astrophysical objects.
As already reported in Section~\ref{sec:2}, the optical counterpart for \four~has been recently studied spectroscopically by \cite{2025PASA...42...85W}.

Only for \three, the angular distance between optical counterparts exceeds 3\arcsec. 
In their note on this source, \cite{2006AJ....131..114B} claimed that their \textit{``identification is uncertain due to a very crowded optical field''}.
We find two GSC\,2.4.2 sources (S9Q4152006 and S9Q4152001) that are closer than S9Q4111853 (the one that we give in Table~\ref{tab:ir_opt}) to the BH06 counterpart (2\arcsec.5 and 2\arcsec.6, respectively, rather than 3\arcsec.2). 
However, these are found at much larger angular distances than S9Q4111853 (3\arcsec.6 and 4\arcsec.0, respectively) from CatWISE J182035.30$-$390928.7.
For these reasons, it is not straightforward to determine which is the GSC\,2.4.2 source corresponding to the BH06 counterpart, but we are led to believe that it is not the one that we give in Table~\ref{tab:ir_opt}.
Thus, we provide a candidate alternative to BH06 for \three, and confirm their counterparts for all the remaining sources.

\begin{table*} 
\begin{center}
\scriptsize
\caption{Comparison between optical counterparts.}
\label{tab:opt}
\begin{tabular}{c|cc|ccc}
\hline
       & \multicolumn{2}{c}{GSC~2.4.2 Source}        &     \multicolumn{3}{c}{\cite{2006AJ....131..114B}}     \\
 Name  & R.A. (J2000) &        Decl. (J2000)         & R.A. (J2000) &          Decl. (J2000)       & $\Delta$ \\
       & ($^{h~m~s}$) & ($^{\circ}$~\arcmin~\arcsec) & ($^{h~m~s}$) & ($^{\circ}$~\arcmin~\arcsec) & (arcsec) \\
 (1)   &     (2)      &              (3)             &      (4)     &             (5)              &   (6)    \\  
\hline                                                                                       
\one   & 17 59 07.13  &        $-$59 46 49.7         & 17 59 07.25  &        $-$59 46 48.9         &   1.23  \\   
\three & 18 20 35.23  &        $-$39 09 29.1         & 18 20 35.48  &        $-$39 09 27.7         &   3.23  \\
\four  & 18 30 58.82  &        $-$36 02 30.4         & 18 30 58.83  &        $-$36 02 31.3         &   1.01  \\
\five  & 20 35 47.68  &        $-$34 54 10.9         & 20 35 47.71  &        $-$34 54 11.5         &   0.65  \\
\hline                                                                                       
\end{tabular}
\end{center}
\tablecomments{\scriptsize{The columns show (1) the name of the SMS4 source; (2) the Right Ascension (J2000) and (3) the Declination (J2000) of the GSC~2.4.2 source that we give as optical counterpart (see Table~\ref{tab:ir_opt}); (4) the Right Ascension (J2000) and (5) the Declination (J2000) of the optical counterpart given by \cite{2006AJ....131..114B} (see their Table~7); (6) the angular separation $\Delta$ with respect to the corresponding GSC~2.4.2 source.\\
}}
\end{table*}

\section{Summary} 
\label{sec:5}
We present a final set of five objects completing our sample of radio sources, selected in \Pa~as having high flux density at 181~MHz and observed since May~2022 as part of a second \sw~observational campaign (PI~Maselli). 
Excluding \four, the other four objects had already been observed but were given as undetected in \Pb, due to short observing times.
The only source in this set of five that was also considered in the X-ray study carried out by \cite{2023ApJS..268...32M} is \five, but their analysis was based on a preliminary set of two observations, for a total exposure of 827~s, and therefore the source was given as undetected.

Following \Pa~and \Pb, for each source, we perform a local source detection after stacking all available observations to derive the intensity and significance of the source (see Section~\ref{sec:3}). 
New observations performed up to February 2026 allow not only the X-ray detection for the first time of \four, but also deeper insight into the X-ray emission of previously undetected sources, providing a detection for \one, \three, and \five.

In contrast, we exclude X-ray emission for \two~with high statistical significance, since the probability that the signal found in 6~ks of exposure is due to a background fluctuation is particularly high ($4.2\cdot10^{-01}$).
A similar situation is found for MRC~B2041$-$604, one of the two sources in Table~3 of \Pb~that was not observed after May 2024, for which $P = 9.9\cdot10^{-02}$ in $\sim$8~ks of exposure.
For the remaining source in Table~3 of \Pb, MRC~B2331$-$416, the possibility of detection with further exposure remains open, since $P = 1.6\cdot10^{-03}$ in 5~ks of exposure.    
Searching in the list of undetected sources in \Pa, we find no more recent than already reported \sw~observations that would allow to update their X-ray information.

For all the \sw~detected sources, in our series of papers, we derive their X-ray properties, including extent and hardness ratio. 
The source extent is determined by computing the ratio of source counts in a circle with radius $\sim12$\arcsec~divided by the source counts in the annulus $\sim24$-48\arcsec, and then comparing this ratio to that expected for an unresolved source, derived from the \sw~PSF (see Section~\ref{sec:3.1} and Table~\ref{tab:er} for details).
As a result, all four X-ray detections presented here are not consistent with point-like sources.
Moreover, hardness ratio analysis (see Section~\ref{sec:3.2} and Table~\ref{tab:hr}) reveals that all sources are mainly characterized by soft X-ray emission, due to the measured $HR$ values.
As a whole, these results can be interpreted in the light of a soft, diffuse X-ray emission surrounding these radio galaxies.
Such emission, that in principle is more difficult to detect than point-like emission, required a substantial increase in exposure, improving the picture reported in \Pb~for these sources.
Much longer exposures, leading to a number of counts high enough to perform a spectral analysis, would be needed to strengthen conclusions for these sources.

For the three sources with available redshift measurements, both spectroscopic and photometric, we compute X-ray unabsorbed flux and luminosity in the 0.3–10 keV band using two different spectral models, both absorbed by the Galactic hydrogen column density, to take into account two different emission mechanisms: a power law with photon index $\Gamma = 2$ for AGN point-like sources and an APEC model (0.4 Solar abundance, $kT = 3$~keV) for thermal emission from a diffuse source.
Luminosities span a range from $\sim 3\cdot10^{42}$ erg~s$^{-1}$ for \four, the only source in our sample with unresolved radio morphology at 5~GHz, relatively close to the observer ($z$=0.078), to $\sim 6\cdot10^{44}$ erg~s$^{-1}$ for \one, an FRII radio galaxy with a photometric redshift that is an order of magnitude higher than \four.

In Section~\ref{sec:3.3}, we compare our results with sources listed in the LSXPS and in the eROSITA-DE DR1 catalogs: using a blind search, \sw~is currently able to detect not only \one, also found in eROSITA-DE DR1, but also \four. 
Furthermore, with our forced photometry, we are able to detect in \sw~data two further sources (\three~and \five) with respect to LSXPS, which basically lists sources detected with a blind search.

In Section~\ref{sec:4}, after verifying the match of the four X-ray detections with the radio sources, we search for infrared and optical counterparts in the AllWISE, CatWISE2020, and the GSC\,2.4.2 catalogs.
As in \Pa~and \Pb, we require a detection in both the infrared and the optical bands to establish a counterpart at lower frequencies for our X-ray detections: as a result, we are able to establish a single candidate counterpart for all the four X-ray detections.

Comparing in Section~\ref{sec:4.3} our results with the counterparts previously proposed by W20 (infrared) and BH06 (optical) our analysis, relying on X-ray source positions, independently confirms the two infrared counterparts provided by W20 for \four~and \five.
Furthermore, we confirm the infrared/optical candidate for \one~previously given as most probable in \Pb: this choice has been recently supported also by \cite{2026MNRAS.tmp..177W}, based on the analysis of radio flux density contours from the Rapid ASKAP (Australian Square Kilometre Array Pathfinder) Continuum Survey \citep{2020PASA...37...48M} at 887.5 MHz (RACS-low1; \citealp{2021PASA...38...58H}).
Finally, we provide a new candidate counterpart for \three: for this source, reported in CatWISE2020 in the infrared band, we also suggest a different counterpart in the optical with respect to BH06.
For the remaining sources, we confirm all the BH06 candidates.
Excluding \four, recently included in a spectroscopic campaign \citep{2025PASA...42...85W}, the remaining optical counterparts still wait to be investigated in greater details.

In conclusion, this third paper of a series completes our investigation, based on two observational campaigns, of the \sw-XRT observations of bright radio sources in the southern hemisphere, comparable to the 3C sample in the north.
As a whole, our analysis leads to the X-ray detection of 30 of the 42 requested targets: five of these X-ray sources, that are due to our forced photometry, are not found in the LSXPS.
Our results support the identification of corresponding counterparts at lower frequencies, highlighting the relevant contribution of the \sw~mission  to the study of these powerful radio sources.

%% Please use the acknowledgment and contribution environments. This will be anonomyized when the "anonymous" style option is used. 
\begin{acknowledgments}
The authors thank the \sw~PI, Brad Cenko, his deputies, and the Science Operations Team for performing the requested observations. 
The authors thank the referee for useful comments and suggestions.
A.M. thanks Silvia Marinoni and Giuseppe Altavilla for fruitful discussions.
W.F., C.J., and R.K. acknowledge support from the Smithsonian Institution and the Chandra High Resolution Camera Project through NASA contract NAS8-03060. W. F. also acknowledges support from NASA Grant GO3-24095X.
A.M. acknowledges financial support from the ASI-INAF agreement No.~2025-33-HH.0. 
This work has been partially supported by the ASI-INAF program I/004/11/4.
This research has made use of archival data, software or online services provided by the ASI Space Science Data Center (SSDC); the High Energy Astrophysics Science Archive Research Center (HEASARC) provided by NASA’s Goddard Space Flight Center; the SIMBAD database and the VizieR catalogue access tool operated at CDS, Strasbourg Astronomical Observatory, France; the NASA/IPAC Extragalactic Database (NED) operated by the Jet Propulsion Laboratory, California Institute of Technology, under contract with the National Aeronautics and Space Administration; the NASA/IPAC Infrared Science Archive, which is funded by the National Aeronautics and Space Administration and operated by the California Institute of Technology. 
\end{acknowledgments}

%\begin{contribution}
%%This section gives authors the space to recognize author contributions. The text inside this environment is NOT counted towards the total word quanta. At a minimum, manuscripts are expected to include this text:

%All authors contributed equally to the Terra Mater collaboration.

%% But authors are expected to provide more specific details, e.g. 
%%
%%SC was responsible for writing and submitting the manuscript.
%%WWM came up with the initial research concept and edited the manuscript.
%%OTS obtained the funding and edited the manuscript.
%%EBF provided the formal analysis and validation. He also edited the manuscript.
%%GEH Supervised the undergraduates, wrote the software and administers the project github and Zenodo repositories.
%%
%% Authors can use the Contributor Role Taxonomy (CRediT) at
%% https://credit.niso.org
%% for ideas on how write a good statement tailored to their needs.

%\end{contribution}

%% To help institutions obtain information on the effectiveness of their telescopes the AAS Journals has created a group of keywords for telescope facilities.
%
%% Following the acknowledgments section, use the following syntax and the
%% \facility{} or \facilities{} macros to list the keywords of facilities used 
%% in the research for the paper.  Each keyword is check against the master 
%% list during copy editing.  Individual instruments can be provided in 
%% parentheses, after the keyword, but they are not verified.
\vspace{5mm}
\facilities{\sw~(XRT), SkyView Virtual Observatory (https://skyview.gsfc.nasa.gov/current/cgi/query.pl), IRSA.}

\bibliography{sms4_III}{}
\bibliographystyle{aasjournal}

%% This command is needed to show the entire author+affiliation list when the collaboration and author truncation commands are used.  
%% It has to go at the end of the manuscript.
%\allauthors

%% Include this line if you are using the \added, \replaced, \deleted commands to see a summary list of all changes at the end of the article.
%\listofchanges

\end{document}